%% file: main.tex
\documentclass[11pt,letterpaper]{article}

\usepackage[letterpaper,margin=1.1in]{geometry}

\usepackage[T1]{fontenc}
\usepackage[utf8]{inputenc}
\usepackage{newtxtext,newtxmath}
\usepackage{microtype}

\usepackage{titlesec}
\usepackage{enumitem}
\usepackage{array}
\usepackage{booktabs}
\usepackage{graphicx}
\usepackage[skip=9pt,labelfont=bf,font=small]{caption}
\usepackage{xcolor}
\usepackage{colortbl}
\usepackage{tikz}
\usetikzlibrary{tikzmark,arrows.meta,calc}
\usepackage{fancyhdr}
\usepackage[hidelinks]{hyperref}
\usepackage{xurl}          % allow long bibliography URLs to break anywhere
\usepackage[backend=biber,style=numeric-comp,sorting=none,
            giveninits=true,maxbibnames=12,doi=false,isbn=false]{biblatex}
\definecolor{fvaccent}{HTML}{B4552D}
\definecolor{fvink}{HTML}{1F1F1D}
\definecolor{fvguar}{HTML}{55534E}
\definecolor{fvmuted}{HTML}{8A8780}
\definecolor{fvrulelight}{HTML}{DCD9D2}
\definecolor{fvrulemid}{HTML}{B4B0A8}

\newcolumntype{L}[1]{>{\raggedright\arraybackslash}p{#1}}

\titleformat{\section}{\large\bfseries}{\thesection}{0.7em}{}
\titleformat{\subsection}{\normalsize\bfseries}{\thesubsection}{0.7em}{}
\titleformat{\subsubsection}{\normalsize\itshape}{\thesubsubsection}{0.7em}{}
\titlespacing*{\section}{0pt}{2.2ex plus .4ex minus .2ex}{1.1ex plus .2ex}
\titlespacing*{\subsection}{0pt}{1.9ex plus .3ex minus .2ex}{0.9ex plus .2ex}
\titlespacing*{\subsubsection}{0pt}{1.7ex plus .3ex minus .2ex}{0.8ex plus .2ex}

\setlist[itemize]{leftmargin=1.5em, itemsep=0.35em, topsep=0.5em, parsep=0pt}

\newcommand{\publicationdate}{September 14, 2026}

\title{\vspace{-3em}\bfseries The Future of Safety for SaMD\vspace{0.4em}}

\author{%
  \normalfont\normalsize
  \begin{minipage}{0.94\textwidth}\centering
    \textbf{Assurea Team}\\[0.15em]
    Rhea Malhotra, \textit{Biomedical Engineering Researcher}
    \;\textperiodcentered\;
    Tanya Sharma, \textit{Co-Founder} \;\textperiodcentered\;
    Krisha Patel, \textit{Co-Founder} \;\textperiodcentered\;
    Satvika Sharma, \textit{Design Verification Researcher}\\[0.7em]
    \textbf{Industry Professionals}\\[0.15em]
    Heena Purkait, \textit{Senior Engineering Project Manager, Fresenius
    Medical Care} \;\textperiodcentered\;
    Mehak Nehal Makhija, \textit{Quality Assurance and Validation Specialist,
    Baxter Healthcare}\\[0.7em]
    \textbf{Runtime Verification Team}\\[0.15em]
    Aellison Cassimiro\textsuperscript{$\ast$}, \textit{Formal Verification Engineer}
    \;\textperiodcentered\;
    Everett Hildenbrandt, \textit{CEO} \;\textperiodcentered\;
    Palina Tolmach, \textit{CTO}\\[0.7em]
    \textbf{Virginia Tech University}\\[0.15em]
    Jaidev Shastri, \textit{Graduate Research Assistant}
  \end{minipage}%
}

\date{\publicationdate}

\begin{document}

\maketitle
\thispagestyle{fancy}

% Corresponding-author note, set as an unnumbered symbol footnote so it does
% not consume a number from the footnote sequence.
\renewcommand{\thefootnote}{\fnsymbol{footnote}}
\footnotetext[1]{Corresponding author: Aellison Cassimiro,
  \href{mailto:aellison.cassimiro@runtimeverification.com}%
       {aellison.cassimiro@runtimeverification.com}.}
\renewcommand{\thefootnote}{\arabic{footnote}}

\begin{abstract}
\noindent
An artificial organ carries failure consequences on the scale of an aircraft
or a reactor, but the software driving it is rarely held to the same standard.
Teams building them rely on testing, which only reaches the failure modes
someone thought of in advance. In a pump or controller that runs inside a
patient for months, the dangerous cases are the ones nobody anticipated.
Formal verification closes that gap. Applied to the device's software, it
proves the code meets its specification for every execution that specification
allows, and where a proof fails, it returns the exact input sequence that
breaks it. The same methods already protect rail, aviation, and nuclear
control systems, and they extend the IEC 62304 lifecycle that a manufacturer
already follows rather than replacing it. In this paper, we explore how to
apply formal verification to artificial organs, stage by stage, and what each
technique actually guarantees about the device.
\end{abstract}

%------------------------------------------------------------------------------
\section{Our Motive}

Within the market for artificial organs, companies must use appropriate means
to verify and validate their software and designs due to the life-critical
nature of their products. Regulatory frameworks like the FDA's and the EU's
Medical Device Regulation ensure that these devices are able to operate
correctly for the sake of a patient's life, and therefore enforce hard audits and guidelines for
validation teams. The current most widely employed method of verification is
the traditional ``spot-checking'', where engineers use tools to test against
specific failure modes in a device's software or design. This approach gives
some assurance as to the efficacy of their product, however, if only certain
genres of failure modes are tested against, then there may be infinite unique
corner cases or other modes that are left unchecked. For this reason,
developers within nuclear, automotive, aerospace, and other industries use an
alternative method for validating their device, one that checks against every
possible failure mode and ensures that their product will always perform as
expected. Formal verification (FV) allows an engineer to say that their
product will at all times perform as outlined by a set of requirements, and
within the medical device world, can give a patient the assurance they need
when dealing with a device so integral to their health and wellness.

%------------------------------------------------------------------------------
\section{What is Formal Verification?}

Let us consider a driver that paces an artificial heart. Its software must
raise an alarm when it detects a fault, hand control to a backup controller if
the main computer fails, and keep pumping while any power source remains. A
test suite can exercise these behaviors for chosen inputs and report what
happened. It tells us whether the software worked for the cases we thought of
running. The behaviors we did not anticipate remain unexamined, and for a
device that runs continuously for months inside a patient, the untested cases
are where the risk is concentrated.

Formal verification takes a different route to confidence by establishing
mathematical arguments that a program satisfies a specification, covering every
execution the specification admits, rather than a sample of them. It works, at
a high level, by translating a device's requirements into a formal model, then
using tools like model checkers and theorem provers to either prove that the
model satisfies those requirements or produce a concrete counterexample showing
exactly where it fails. That counterexample is the real value: instead of a
vague sense that something might be wrong, engineers get an exact input
sequence or state that breaks the system.

Within a given device, one that possesses software that drives the function of
the product, certain properties of this software can be formally specified. If
a property reflects a behavior that must be truthful at all times of the
system execution, we can refer to it as an invariant. Written in logic rather
than prose, such statements become objects that a tool can verify. These
invariants can be tested to some extent using the approaches currently used in
the field, including unit tests, integration tests, coverage testing, fuzzing,
and other techniques. Still, these do not provide concrete proof with
certainty of safety. A formally verified property is guaranteed to hold for
any configuration that the system can find itself in. In other words, while
traditional testing approaches guarantee that a device's functionality works
for a specific input or set of inputs, formal verification can validate these
same functionalities for all semantically valid inputs, either mathematically
proving safety or identifying configurations in which the device's safety is
compromised.

Work on proving properties of programs began in the 1960s and has been going
on continuously since. The logics came first, then machine assistance for the
proofs, then solvers fast enough for industrial problems, and more recently
analysers that work on production source. The earliest safety cases built on
proof date from around 1990, and systems verified this way have entered
service ever since, several still running today. A team adopting these methods
now takes up a discipline developed and tested in the field for more than
sixty years.

Formal verification as a field is recognized for the strength of the guarantees
it provides, and it is used in multiple crucial systems with the same level of
severity as medical devices. Some of them are:

\begin{itemize}
  \item Rail (Paris Métro): The driverless trains on Line 14 run on software
        whose logic was proven mathematically before it was written as code.
        Proof replaced unit testing entirely, and the developers report no
        faults found since the line opened in October
        1998.~\cite{abrial1999meteor,clearsy2024line14}

  \item Air travel (Airbus flight controls): The software moving the control
        surfaces on the A340 and A380 was analyzed across every possible
        execution, not a selection of test cases, and shown to be free of
        arithmetic and memory faults even in its floating-point
        calculations.~\cite{souyris2007astree,astree}

  \item Naval aviation (helicopter landing advice): A system advising crews
        whether a helicopter can safely land on a moving warship was built for
        proof, with roughly 150 proofs on its specification and 9,000 on its
        code. The team found proof was more efficient at catching faults than
        the testing they ran alongside it.~\cite{king2000proof}

  \item Nuclear power (reactor shutdown): The software shutting down the
        reactors at Ontario's Darlington station was written as precise tables
        of required behavior, with a theorem prover confirming the design met
        them. It stands among the earliest cases of a national regulator
        accepting mathematical proof in place of testing
        alone.~\cite{lawford_nuclear}

  \item Spacecraft (NASA probe): Before Deep Space 1 launched, formal analysis
        of its autonomous control software found five timing faults that the
        developers said testing would have missed. Faults of that same kind
        later suspended the probe's operations in
        flight.~\cite{havelund2001spin,havelund_remoteagent}
\end{itemize}

%------------------------------------------------------------------------------
\section{Closing the Verification Gap}

Artificial organs carry the failure consequences of aerospace or nuclear
control systems, but rarely the verification rigor of those industries. Formal
verification (FV) closes that gap with exhaustive mathematical proof that a
system behaves correctly across every possible input and state instead of just
the scenarios an engineer tests. Before any artificial organ reaches a
patient, it has to clear regulatory approval, whether that's the FDA in the
United States, the EU's Medical Device Regulation in Europe, or an equivalent
regime elsewhere in the world. Despite operating in different countries, these
regulators largely converge on the same underlying standards for medical
device software, requiring manufacturers to prove their devices have been
thoroughly checked for errors before reaching patients. Right now, most
companies do this through testing: running the device through thousands of
scenarios and checking that nothing goes wrong. Formal verification covers
those scenarios and every single other one, leaving nothing to chance and
ensuring successful operation. For a device like an artificial heart or an
insulin pump, where a single overlooked scenario could be fatal, that
distinction matters. Because formal methods produce a clear, provable record
of what was checked and why, they fit naturally into the kind of documentation
regulators worldwide already expect, which makes them not just safer, but
easier for regulators to trust~\cite{fda_infusionpump}, regardless
of where the device is headed to market. To assess and analyze how that
integration would occur, and its benefits within the company and for
consumers, let us examine the pathway for incorporating formal methods into
the existing device development.

%..............................................................................
\subsection{Incorporating FV Methodologies into Artificial Organs \& Why It Is
an Effective Solution}

While artificial organs are a relatively novel innovation, it is clear that
their need for safety is paramount. Any small infraction of the device
requirements or a system bug can snowball into catastrophic failure, resulting
in medical complications or even death. Medical devices with software-based
systems benefit from the coded formal model, while mechanical,
software-lacking devices have formalized continuous properties that address
mechanical failure. Where traditional testing leaves gaps in assurance through
unexplored corner cases, timer overflows, and other limitations, formal methods
ensure that the device is working correctly as defined at all times.

The development of medical device software, as laid out by IEC
62304~\cite{iec62304}, begins by classifying the software according to how much
harm its failure could cause, which sets how demanding every later step
becomes. The team then fixes its ground rules, naming the coding standards,
methods and tools it will use. Engineers verify each unit of code against
acceptance criteria they have defined, usually by testing and measuring how
much of the code those tests reach. They then combine the units and check that
the interfaces behave and that the safety measures engage, often by
deliberately inducing faults. The assembled software runs on the real hardware.
At release, every known defect still present is recorded with a written
justification for leaving it. Once the device reaches patients, later changes
are handled as maintenance, with each change assessed for the risk it
introduces and the untouched software re-tested around it. Version control and
problem reporting run throughout. Every step in this sequence rests on running
the software and observing what happens, which means the evidence it produces
covers the situations the engineers thought to create.

The design stage is the primary and most crucial point for FV incorporation;
however, the methodology can be applied throughout the development lifecycle.
The differences between the two main device types (software-based vs purely
mechanical) lie primarily in how formal methods are approached, rather than
directly employed. A device's properties are outlined similarly, and hazards
are established. The divergence occurs when setting physical versus software
requirements, and the following models are generated and tested. Once
properties are proved and counterexamples or violations are fixed, device
compliance regulation submissions can occur (FDA, CE Mark, etc.).

FV's most mature tooling, including model checkers and theorem provers, target
software and digital logic. For software-driven organs (insulin pumps,
pacemaker firmware, dialysis
controllers)~\cite{arney2007gip,zhang2011insulin,masci2014pvs}, the path is
borrowing directly from aerospace and automotive safety-critical
practice~\cite{do178c,rtca_do333,iso26262}.

For primarily mechanical or biomaterial devices (a mechanical heart valve, a
biomaterial scaffold), FV's role narrows to the control systems, design, build,
and safety interlocks around the device, which ensure the device's physical
operation. Formal methods are applicable across all medical devices, and the steps
taken to ensure the formal properties shift accordingly. It is important to
highlight that, while it is possible and extremely beneficial to apply FV in
the development of purely mechanical medical devices, the process of applying
formal methods to it becomes something that is tailored.

\input{figures/fv-lifecycle-figure}

As seen in Figure~\ref{fig:pathway}, formal methods change what each of those
checks establishes. Each phase has an added security technique using applied
formal methods to generate artefacts. Artefacts are any concrete work product
the process creates and keeps. It can be a document, a model, a specification,
a set of test cases, or a proof that represents the goal of that stage.

Planning gains two products the process did not have before: the properties
themselves, and a register of the assumptions they depend on. At the
\emph{units} stage, sound static analysis~\cite{cousot1977abstract} reasons
over every state the program could reach, without running any of it, and
proves the code cannot overflow, divide by zero or read outside an array on
any run. It needs nothing specified in advance. Deductive verification goes
further. A contract on each unit states what it needs from its caller and what
it promises in return~\cite{hoare1969axiomatic,meyer1992contract}, tools
translate the code and its contract into logical obligations, and a prover
discharges them. Each proof holds for every input the contract permits, and it
applies to the shipping source. The question shifts from whether the chosen
test values behaved correctly to whether any value could behave otherwise.
During \emph{integration}, model
checking~\cite{clarke1981design,queille1982cesar} takes a simplified model of
the design, records the states the system can occupy and how it moves between
them, and explores every path the model allows. Claims about failover and
deadlock get settled across all interleavings rather than the handful of
faults an engineer thought to induce, and a failing property comes back as a
step-by-step trace to the violation. At the \emph{system} stage, the same
exploration, extended over time, runs against a model of the patient and pump,
and bounds the worst-case for alarm and pump-cycle deadlines instead of
reporting the slowest case seen on the bench. At \emph{release}, the record
lists which properties were proven and which remain open, stating residual
risk in terms of behavior rather than a count of known defects. During
maintenance, the team re-establishes only the proofs that a change breaks, so
later versions inherit proofs instead of assumptions. Once the device is in
service, runtime verification compiles those same properties into monitors
that run beside the software and report a violation the moment it occurs.

These are just a few examples of techniques available to enhance the security
of the medical device development lifecycle.

\subsubsection{Industry Perspective by Heena Purkait: Engineering Project
Management}

From a project management perspective, the value of formal methods within an
artificial organ development process is closely tied to how regulatory
expectations shape the schedule itself. Regulators may converge on the same
underlying standards, but compliance execution rarely does. Working across
EMEA, every country carries its own certification requirements and its own
logic, and obligations like GDPR can stall an otherwise ready product if they
aren't accounted for at the design stage. That gap between shared standards and
fragmented compliance is where a provable record matters most: each company's
QMS runs on SOPs built around FDA guidance, and a formal, traceable model of
what was checked and why holds up the same across every regulator's
submission, even when the surrounding paperwork does not. Verification rigor
does not stay uniform across a project either, it shifts with device complexity
and with where the product is being submitted. If the right people are not
assigned early, or a defect turns out to be a requirements gap rather than an
implementation error, that is a coordination failure as much as a technical
one. Formal verification, integrated early and tied to SMART, specific,
measurable, achievable, relevant, time-bound goals, gives a way to surface
those gaps before they collapse a timeline late in the project.

Selling that shift to stakeholders who are measured on delivery dates, not
defect rates, is the harder part of the role. It comes down to making sure
every function (engineering, quality, human factors, and risk) is informed at
each phase, and reminding people of their responsibility when it is easy to
lose track in a complex project. Existing methodologies like RACI (responsible,
accountable, consulted, informed) and each company's QMS systems can be better
supported by formal verification methods, and when these engines are able to
produce definable traces to the root causes of a bug, then each team member
benefits from that clarity.

Ultimately, the integration of FV within the standard artificial organ
development pathway occurs at each and every stage, and every team member is
responsible for engaging with formal methodologies. This process allows us to
identify risks and errors early on and move forward efficiently, and as a
project manager, this methodology can be integrated seamlessly within our
scheduling and regulatory standard teams, which are better equipped to deal
with issues.

%..............................................................................
\subsection{Quality Considerations for Employing Formal Methods}

Formal Verification is an addition to existing Quality Assurance (QA)
processes, not a replacement. Manufacturers in the US, for example, operate
under ISO 13485~\cite{iso13485}, IEC 62304, and FDA design
controls. FV slots into that framework as a rigor layer, not a replacement for
it.

The necessary scaffolding for the generation of proofs over properties can be
abstracted to a two-stage process. First, someone must write down what
``correct'' means with enough precision to reason about. Second, a tool must
relate these statements to the program.

The first stage encompasses the definition of a set of properties, which are
formally expressed claims about the software's behavior. As previously
explored, we can also mature our properties into invariants once we define the
behaviors that should hold at any given moment in the device's lifecycle.

The second element varies by technique, which can be used in conjunction to
identify issues both related and unrelated to the properties and invariants
raised so far. These invariants are the guiding principles of the strategies
employed during the development process, explored in Figure~\ref{fig:pathway},
and it is of utmost importance that they are well defined and understood. Both
failure and correct functioning must be clear at design time to assure the
highest extraction of value possible from the formal verification techniques.

\subsubsection{Industry Perspective by Mehak Makhija: QA \& Validation}

From a Quality Assurance and Validation perspective, Formal Verification (FV)
should strengthen, not replace, traditional verification and validation methods
by addressing the finite limitations of current testing. It shifts the focus
from checking representative scenarios to mathematically proving that critical
safety properties hold across all possible system states. This is particularly
valuable because software failures can emerge from rare interactions between
functions that individually passed verification. In traditional QA practice,
these gaps may be identified proactively through FMEAs, risk
assessments~\cite{iso14971}, design and protocol reviews, or traceability
assessments; others may only surface through deviations, nonconformances,
complaints, or post-market investigations. An intermittent timing or
communication failure, for example, may require event-log reconstruction,
root-cause analysis, corrective action, and expanded regression testing before
the failure sequence is understood. Applying formal methods earlier creates an
opportunity to identify some of these edge cases before they become validation
failures, CAPAs, field corrections, or patient safety events.

From a validation standpoint, FV can be integrated directly into the existing
requirements-to-risk-to-verification traceability structure. A safety-critical
requirement, such as an artificial pancreas suspending insulin delivery under
an unsafe glucose condition or an artificial heart controller maintaining safe
operation during failover, would still link to its hazard analysis, risk
controls, verification activities, acceptance criteria, and validation
evidence. Formal proof simply adds another layer of objective evidence for
design review and regulatory assessment. However, FV also introduces
responsibilities for defining appropriate safety properties, validating
assumptions, maintaining models as designs change, and interpreting proof
results. A proof is only as meaningful as the requirements and model on which
it is based, making QA oversight and traceability essential. Ultimately, FV's
greatest value is defect prevention. Stronger audit defensibility and
regulatory readiness are important benefits, but for artificial organs and
other life-sustaining devices, the primary objective is to identify critical
failures as early as possible and prevent them from reaching the patient.

%..............................................................................
\subsection{Traditional Verification's Gaps within Life-Critical Environments}

Within an artificial organ, the situation is as immediately life-critical as
aerospace or nuclear environments, if not more so. A single bug could cause
immediate health issues or death (cardiac arrest, glucose overdose, etc).
Formal verification closes the gap that traditional spot testing leaves open,
oftentimes missing race conditions at exact clock boundaries, mode-transition
deadlocks (two valid moves collide), timer overflows, threshold boundary bugs
(sensor exactly at limit), general concurrency and non-determinism issues
between different modules of a complex system, and unexplored corner cases. It
is not that traditional approaches neglect these properties, but rather that
they have limitations on how to test them. Integrating formal
verification into the regular development pathway for an artificial organ will
ultimately increase the safety coverage of the claim a medical device company
can make. A checkable mathematical object is categorically stronger than ``X
tests passed at Y\% coverage'' and directly strengthens the FDA Class III
safety-and-effectiveness case.

The effectiveness of this methodology lies in its ability to ensure a device
that will never compromise a patient's life: an artificial organ that cannot
spontaneously reach a shutdown state and cause immediate organ failure, or
encounter a unique corner failure mode and enter a static or false state that
administers the incorrect amount of medication. This level of assurance is not
only superior to that of traditional testing, but vital in the life-critical
environment that medical devices such as artificial organs exist within.

Testing samples the input space, while a proof ranges over everything the
specification admits, which makes coverage measurement largely beside the point
at the unit level. The guarantee is complete in one direction and conditional
in another. It holds for every input and every execution. Although it only
holds for the properties someone wrote down, under the assumptions someone
recorded, and, when a model is used, only as far as that model matches the
code. The gaps, therefore, do not disappear. They move to three places that
can be named and managed: properties nobody thought to state, assumptions
about hardware and environment taken on trust, and the argument connecting a
model to the software it stands for. Runtime monitoring addresses the second of
these by watching in service for assumptions that could not be settled. The
practical difference is that what remains uncertain becomes explicit. In a
testing-only process, whatever was not examined stays unknown and uncounted.
With proofs in place, whatever remains unproven has a name, can be written into
the safety file, and can be argued about or watched.

%------------------------------------------------------------------------------
\section{To Conclude}

A patient-oriented market such as that of artificial organs requires certain
levels of trust between both the company and the person utilizing their
product. That trust ensures the safety and health of the patient, and it is
dependent on the regulations set by the governing organization and the
validation methods of the company. If this trust can be fortified and the
device's design verified against each and every possible failure mode, then
customers would be more willing to reach out to companies in pursuit of a
better quality of life facilitated by their products. Formal verification
methods are the key to strengthening trust, and employing methods like model
checking, timed automata, and more can allow a safer and more trustworthy
environment within the patient community.

%------------------------------------------------------------------------------
\printbibliography[title={References}]

\end{document}

%% file: figures/fv-lifecycle-figure.tex
%==============================================================================
%  Figure 1 --- Formal Verification Applied to the Artificial Organ Software Development Pathway
%
%  Set natively in LaTeX as a tabular with a TikZ overlay, so the type sets at
%  the document's own size and stays searchable and selectable.
%
%  Needs at least two passes: the process arrow and phase dots are placed with
%  TikZ \tikzmark, which resolves on the second run.
%==============================================================================
\begin{figure}[!htb]
\centering
\begingroup
%---- local typographic settings ---------------------------------------------
\sffamily\scriptsize
\setlength{\tabcolsep}{3pt}
\setlength{\aboverulesep}{2.5pt}
\setlength{\belowrulesep}{2.5pt}
\renewcommand{\arraystretch}{1.12}
%---- local cell styles -------------------------------------------------------
\newcommand{\fvhdr}[1]{{\tiny\bfseries\color{fvmuted}#1}}
\newcommand{\fvstage}[1]{{\footnotesize\bfseries\color{fvink}#1}}
\newcommand{\fvclause}[1]{{\color{fvaccent}#1}}
\newcommand{\fvact}[1]{{\color{fvink}#1}}
\newcommand{\fvactm}[1]{{\color{fvmuted}\itshape #1}}
\newcommand{\fvguar}[1]{{\color{fvguar}#1}}
\noindent\begin{minipage}{\textwidth}
\begin{tabular}{@{}L{24pt}L{45pt}L{56pt}L{164pt}L{138pt}@{}}
\arrayrulecolor{fvrulemid}\specialrule{0.7pt}{0pt}{3pt}
 & & & \fvhdr{ARTEFACT BUILT \textperiodcentered\ CHECK PERFORMED}
     & \fvhdr{GUARANTEE OBTAINED} \\[2.5pt]
\tikzmark{g1} & \fvstage{Classify} & \fvclause{Cl.~4.3}
  & \fvactm{No new artefact --- classification fixes how much must be proven}
  & \fvguar{Depth of formal evidence follows the safety class} \\
\arrayrulecolor{fvrulelight}\cmidrule[0.45pt]{2-5}
\tikzmark{g2} & \fvstage{Plan} & \fvclause{Cl.~5.1.4 \textbf{(C)}}
  & \fvact{\textbf{Write} the property set and assumption register (Cl.~5.2);
           \textbf{build} the design model (Cl.~5.3); select and
           \textbf{qualify} the tools}
  & \fvguar{A checkable definition of correct behaviour, and tools whose
            results may be relied on as evidence} \\
\arrayrulecolor{fvrulelight}\cmidrule[0.45pt]{2-5}
\tikzmark{g3} & \fvstage{Units} & \fvclause{Cl.~5.5.2--5.5.5 \textbf{(C)}}
  & \fvact{\textbf{Derive} unit contracts from the property set and the
           decomposition (Cl.~5.4); \textbf{write} them into the source and
           discharge by proof; sound static analysis;
           \textit{unit tests remain}}
  & \fvguar{Correct for all inputs, not a sampled subset, and free of runtime
            errors. Proven on the source, so no model-to-code gap arises
            here} \\
\arrayrulecolor{fvrulelight}\cmidrule[0.45pt]{2-5}
\tikzmark{g4} & \fvstage{Integration} & \fvclause{Cl.~5.6, 7.3}
  & \fvact{Model-check the design model; \textbf{argue} that composed unit
           contracts entail it, and that the model abstracts the code}
  & \fvguar{Every interleaving explored, and unit proofs made to bear on
            system properties. Subject to both arguments holding} \\
\arrayrulecolor{fvrulelight}\cmidrule[0.45pt]{2-5}
\tikzmark{g5} & \fvstage{System} & \fvclause{Cl.~5.7}
  & \fvact{\textbf{Build} a plant and patient model; verify timing and
           closed-loop behaviour against it}
  & \fvguar{Worst-case deadlines bounded and the loop shown to stay in a safe
            envelope, given stated assumptions} \\
\arrayrulecolor{fvrulelight}\cmidrule[0.45pt]{2-5}
\tikzmark{g6} & \fvstage{Release} & \fvclause{Cl.~5.8}
  & \fvact{\textbf{Assemble} the evidence into an assurance case; carry open
           properties and unmet assumptions into the risk file}
  & \fvguar{Residual risk stated as named unproven properties, not only as a
            count of known defects} \\
\arrayrulecolor{fvrulelight}\cmidrule[0.45pt]{2-5}
\tikzmark{g7} & \fvstage{Change} & \fvclause{Cl.~6.2, 7.4}
  & \fvact{\textbf{Update} property set, design model and contracts together;
           re-establish only the proofs the change invalidates}
  & \fvguar{A revision inherits proofs rather than assumptions, with no
            accumulating unverified legacy} \\
\arrayrulecolor{fvrulemid}\specialrule{0.7pt}{3pt}{3pt}
\tikzmark{g8}\fvactm{spans all} & \fvstage{Always} & \fvclause{Cl.~8, 9}
  & \fvact{\textbf{Version} properties, model, contracts and proofs with the
           code; \textbf{monitor} at runtime the assumptions proof left
           undischarged}
  & \fvguar{Evidence stays synchronised with the code, and undischarged
            assumptions are watched in service} \\
\arrayrulecolor{fvrulemid}\specialrule{0.7pt}{3pt}{0pt}
\end{tabular}%
%
%---- gutter graphics: process arrow, phase dots, change-to-plan return edge --
\begin{tikzpicture}[remember picture, overlay,
                    dot/.style={circle, inner sep=0pt, minimum size=4.4pt}]
  % Spine and return edge share the gutter column.  \gx is the spine's offset
  % from the gutter's left edge, \gr the return edge's; the 14pt between them
  % is what makes the return read as a bracket rather than a parallel line.
  % Both sit inside the 24pt gutter, so no column width changes and no reflow.
  \def\gx{18pt}
  \def\gr{4pt}
  % the spine runs from above the Classify dot to the rule below Change
  \draw[fvrulemid, line width=1.1pt, -{Stealth[length=4pt, width=3.4pt]}]
        ([xshift=\gx, yshift=7pt]pic cs:g1)
     -- ($([xshift=\gx]pic cs:g7)!0.62!([xshift=\gx]pic cs:g8)$);
  % Dashed return edge: a modification re-enters the lifecycle at Plan.  Drawn
  % as a rectangular bracket -- out from the spine below Change, up the gutter,
  % back in to the Plan dot -- with rounded corners.
  \draw[fvaccent, line width=0.8pt, dash pattern=on 2pt off 1.6pt,
        rounded corners=3pt, -{Stealth[length=4pt, width=3.4pt]}]
        ([xshift=\gx,  yshift=-9pt]pic cs:g7)
     -- ([xshift=\gr,  yshift=-9pt]pic cs:g7)
     -- ([xshift=\gr,  yshift=2.2pt]pic cs:g2)
     -- ([xshift=15pt, yshift=2.2pt]pic cs:g2);
  % phase dots: hollow for every stage on the spine, filled for the re-entry point
  \foreach \i in {1,3,4,5,6,7}{%
    \node[dot, fill=white, draw=fvrulemid, line width=0.8pt]
         at ([xshift=\gx, yshift=2.2pt]pic cs:g\i) {};}
  \node[dot, fill=fvaccent, minimum size=5pt]
       at ([xshift=\gx, yshift=2.2pt]pic cs:g2) {};
\end{tikzpicture}%
\end{minipage}
\endgroup
\caption{Formal Verification Applied to the Artificial Organ Software Development Pathway}
\label{fig:pathway}
\end{figure}